\documentclass{ws-rv975x65}
\usepackage{subfigure}     
\usepackage{ws-rv-thm}     
\usepackage{ws-rv-van}     
\makeindex
\def\prd{PRD}
\def\jcap{JCAP}
\def\apj{ApJ}
\def\apjl{ApJ}
\def\aap{A\&A}
\def\mnras{MNRAS}

\begin{document}

\chapter[The origin of IceCube neutrinos]{The origin of IceCube's neutrinos:\\
Cosmic ray accelerators embedded in star forming calorimeters}\label{W-chapter}

\author[E. Waxman]{E. Waxman
}

\address{Particle physics and Astrophysics department,\\ Weizmann Inst. of Science, Rehovot 76100, Israel
}

\begin{abstract}
The IceCube collaboration reports a detection of extra-terrestrial neutrinos. The isotropy and flavor content of the signal, and the coincidence, within current uncertainties, of the 50~TeV to 2~PeV flux and the spectrum with the Waxman-Bahcall bound, suggest a cosmological origin of the neutrinos, related to the sources of ultra-high energy, $>10^{10}$~GeV, cosmic-rays (UHECR). The most natural explanation of the UHECR and neutrino signals is that both are produced by the same population of cosmological sources, producing CRs (likely protons) at a similar rate, $E^2d\dot{n}/dE\propto E^{0}$, over the [1~PeV,$10^{11}$~GeV] energy range, and residing in "calorimetric" environments, like galaxies with high star formation rate, in which $E/Z<100$~PeV CRs lose much of their energy to pion production.

A tenfold increase in the effective mass of the detector at $\gtrsim100$~TeV is required in order to significantly improve the accuracy of current measurements, to enable the detection of a few bright nearby starburst "calorimeters", and to open the possibility of identifying the CR sources embedded within the calorimeters, by associating neutrinos with photons accompanying transient events responsible for their generation. Source identification and a large neutrino sample may enable one to use astrophysical neutrinos to constrain new physics models.

\end{abstract}
\body

\section{Introduction}\label{sec:intro}

The sources and acceleration mechanism of cosmic-rays of different energies have not been reliably identified despite many decades of research\cite{Helder12_SNR_CR,Lemoine13}. Particularly challenging to models are the observations of UHECRs, since most models cannot account for the highest observed particle energies\cite{W11_frontiers_review,Lemoine13}. One of the main goals of the construction of high energy neutrino telescopes is to resolve the open questions associated with these long standing puzzles\cite{Gaisser95_PhysRep,W11_frontiers_review}.

Assuming that UHECRs are charged nuclei accelerated electromagnetically to high energy in astrophysical objects, some fraction of their energy is expected to be converted to high energy neutrinos through the decay of charged pions produced by the interaction of cosmic-ray protons/nuclei with ambient gas and radiation. A detection of this neutrino signal may enable the identification of the sources and will provide qualitative new constraints on accelerator models. The upper bound derived by Waxman \& Bahcall (WB) on the neutrino intensity produced by the CR sources\cite{WBbound1,WBbound2} implies that a giga-ton neutrino telescope is required to detect the expected flux in the energy range of $\sim1$~TeV to $\sim1$~PeV, and that a much larger effective mass is required at higher energy (see fig.~\ref{fig:WBbound}).

In this chapter we explain the reasoning leading to the conclusion, that the extra-terrestrial flux of neutrinos detected by the giga-ton IceCube detector\cite{IC14PhRvL_3yr_detection,IC15_nu_mu_detection} is produced by UHECR sources embedded in "calorimetric" environments, characteristic of the conditions in galaxies with high star formation rate.
The derivation of the WB bound is described in \S~\ref{sec:WB}. The implications of IceCube's detection and the constraints it imposes on the sources of high energy neutrinos and CRs are described in \S~\ref{sec:IC_nu_origin}. The main conclusions, the main open questions and the prospects for progress in the study of CR sources
using high energy neutrinos of astrophysical origin are described in \S~\ref{sec:discussion}.

\section{The Waxman-Bahcall bound}\label{sec:WB}

\subsection{UHECR composition, production rate and spectrum}\label{sec:UHECRs}

The composition of UHECRs is controversial, with air-shower data from the Fly's Eye, HiRes and Telescope Array observatories\cite{Bird93_FE_comp,HiRes_Composition_05,TA_Comp_2015} suggesting a proton dominated composition and data from the Pierre Auger Observatory\cite{Auger_comp_2014} suggesting a transition to heavier elements above $10^{10}$~GeV. Due to this discrepancy, and due to the experimental and theoretical uncertainties in the relevant high energy particle interaction cross sections used for modeling the shape of the air showers, it is impossible to draw a definite conclusion regarding composition based on air-shower data at this time (the anisotropy signal provides an indication for a proton dominated composition\cite{LW09_Aniso_Comp}, but is so far detected with only a $\sim2\sigma$ confidence level\cite{Lemoine13}).

The observed flux and spectrum of $E>10^{10.2}$~GeV CRs is consistent with a cosmological distribution of cosmic-ray sources, producing protons at a rate\cite{W95_QUHECR,BW03_QUHECR,Katz_UHECR_09}
\begin{equation}\label{eq:uhecr}
Q_{\rm UHE}\equiv\left(E_p^2d\dot{n}_p/dE_p\right)_{z=0} = 0.5\pm0.15\times10^{44} {\rm erg/Mpc^{3}yr}.
\end{equation}
The energy density of CRs (at different energies) is determined by the (energy dependent) CR production rate and energy loss time. The energy loss of protons is dominated at high energy by the production of pions in interaction with cosmic microwave background (CMB) photons\cite{Greisen,ZK}. The rate estimate of eq.~\ref{eq:uhecr} is based on the direct measurement of UHECRs and on the well understood physics of proton-CMB interaction. It is therefore accurate (to $\sim 30\%$) as long as the composition is dominated by protons. Observations determine only the local, $z=0$, proton production rate and spectrum, since the energy loss time of protons is much smaller than the Hubble time (e.g., $\sim3\times10^8$~yr corresponding to a propagation distance of $\sim100$~Mpc at $10^{11}$~GeV).

The observed UHECR spectrum is consistent with a "flat" proton generation spectrum, $d\log\dot{n}/d\log E\approx -2$ with equal energy produced per logarithmic CR energy interval, modified by interaction with CMB photons. This supports a proton dominated composition, since a flat generation spectrum is observed in a wide range of astrophysical environments (e.g. CR protons in the Galaxy\cite{Blandford87,Axford94}, electrons in supernova remnants\cite{Blandford87,Axford94} and in $\gamma$-ray bursts\cite{Waxman_rel-plasma_rev06}), and is a robust prediction of the best understood and most widely accepted model for particle acceleration in astrophysical objects- Fermi acceleration in collisionless shocks\cite{Blandford87,Bednarz98RelFermi,Keshet05RelFermi} (although a first principles understanding of the process is not yet available).

If the composition is dominated by heavier nuclei up to iron (e.g. O, Si, Fe), the inferred energy generation rate at $10^{10.5}$~GeV would change by a factor of only a few. This is due to the fact that the energy loss distance of protons is not very different from that of heavy nuclei (due to photo-disintegration interactions with the infra-red background)\cite{Allard12}. However, the different energy dependence of the energy loss distances of heavy nuclei and protons, implies that for a heavy nuclei composition, a generation spectrum different from $E^2d\dot{n}/dE\propto E^0$ or an ad-hoc energy dependent composition would be required to fit the observed spectrum\cite{Allard12}. We consider this as further evidence supporting the proton dominated flux hypothesis, which we adopt in what follows. We return to the possibility of heavy nuclei domination in \S~\ref{sec:calorimeters}.

\subsection{The neutrino intensity limit}\label{sec:WBder}

The energy production rate, eq.~\ref{eq:uhecr}, sets an upper bound to the neutrino intensity produced by sources, which are optically thin for high-energy nucleons to $p\gamma$ and $pp(n)$ interactions. For sources of this type, the energy generation rate of neutrinos can not exceed the energy generation rate implied by assuming that all the energy injected as high-energy protons (eq.~\ref{eq:uhecr}) is converted to pions (via $p\gamma$ and $pp(n)$ interactions). The resulting all-flavor upper bound is\cite{WBbound1,WBbound2}
\begin{equation}
E_\nu^2\Phi_{\rm WB,\, all\, flavor}=3.4\times10^{-8}\frac{\xi_z}{3}\left[\frac{(E_p^2d\dot{n}_p/dE_p)_{z=0}}{0.5\times10^{44}{\rm erg/Mpc^3yr}}\right]{\rm GeV/cm^{2}s\,sr},
\label{eq:WB}
\end{equation}
where $\xi_z$ is (a dimensionless parameter) of order unity, which depends on the redshift evolution of $E_p^2d\dot{n}_p/dE_p$. The value $\xi_z=3$ is obtained for rapid redshift evolution, $\Phi(z)=(1+z)^3$ up to $z=2$ and constant at higher $z$, corresponding approximately to that of the star-formation rate or AGN luminosity density evolution ($\xi_z=0.6$ for no evolution). The numerical value (3.4) given in eq.~\ref{eq:uhecr} is obtained for equal production of charged and neutral pions, as would be the case for $p\gamma$ interactions dominated by the $\Delta$ resonance\cite{WBbound1}. For $p\gamma$ interactions at higher energy, or $pp(n)$ interactions, the charged to neutral pion ratio may be closer to 2:1, increasing the bound flux by $\approx30\%$.

\begin{figure}[ht]
\centerline{
  \subfigure[The WB bound]
     {\includegraphics[width=2.5in]{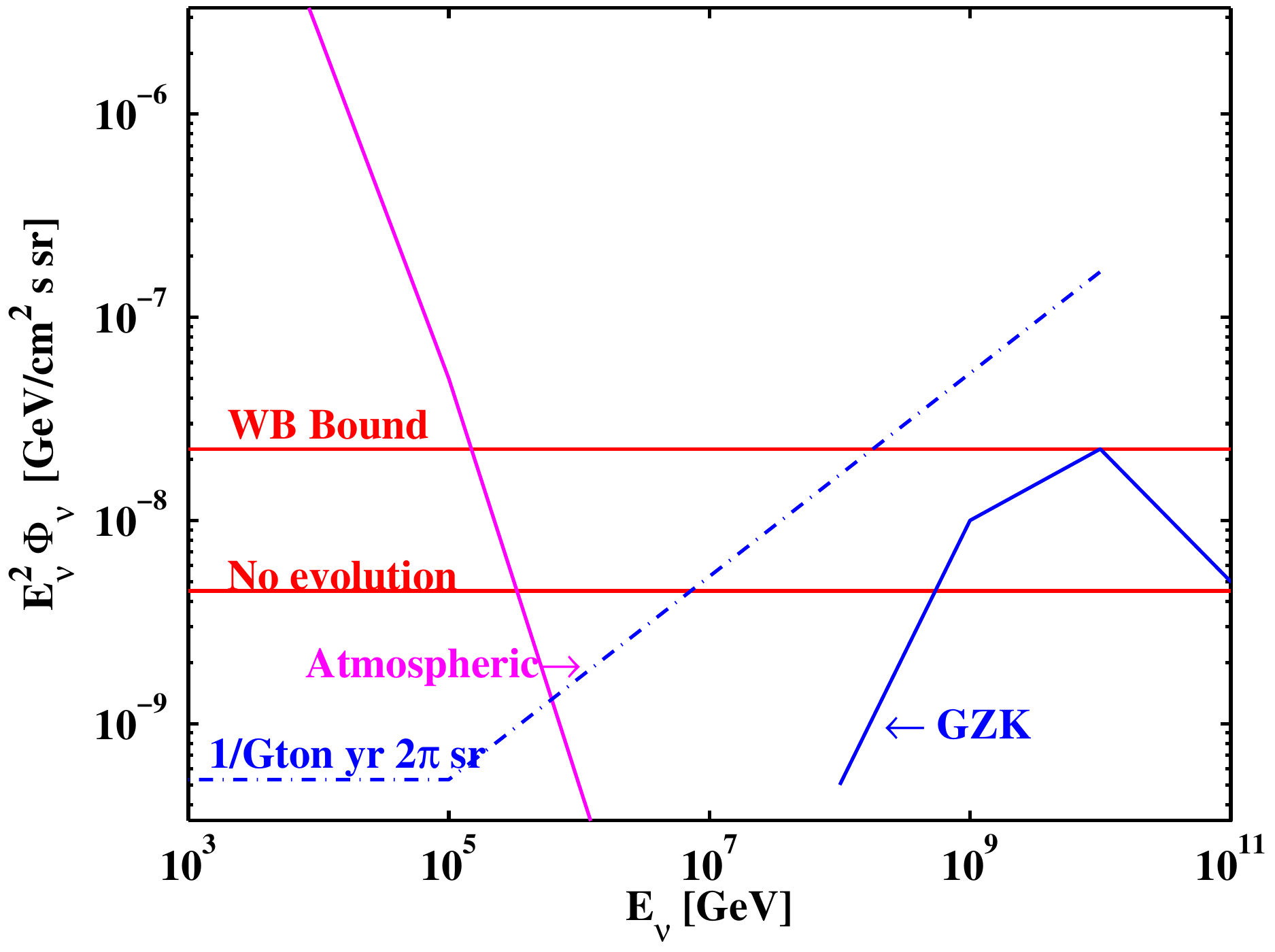}\label{fig:WBbound}}
  \hspace*{4pt}
  \subfigure[IceCube's detection]
     {\includegraphics[width=2.5in]{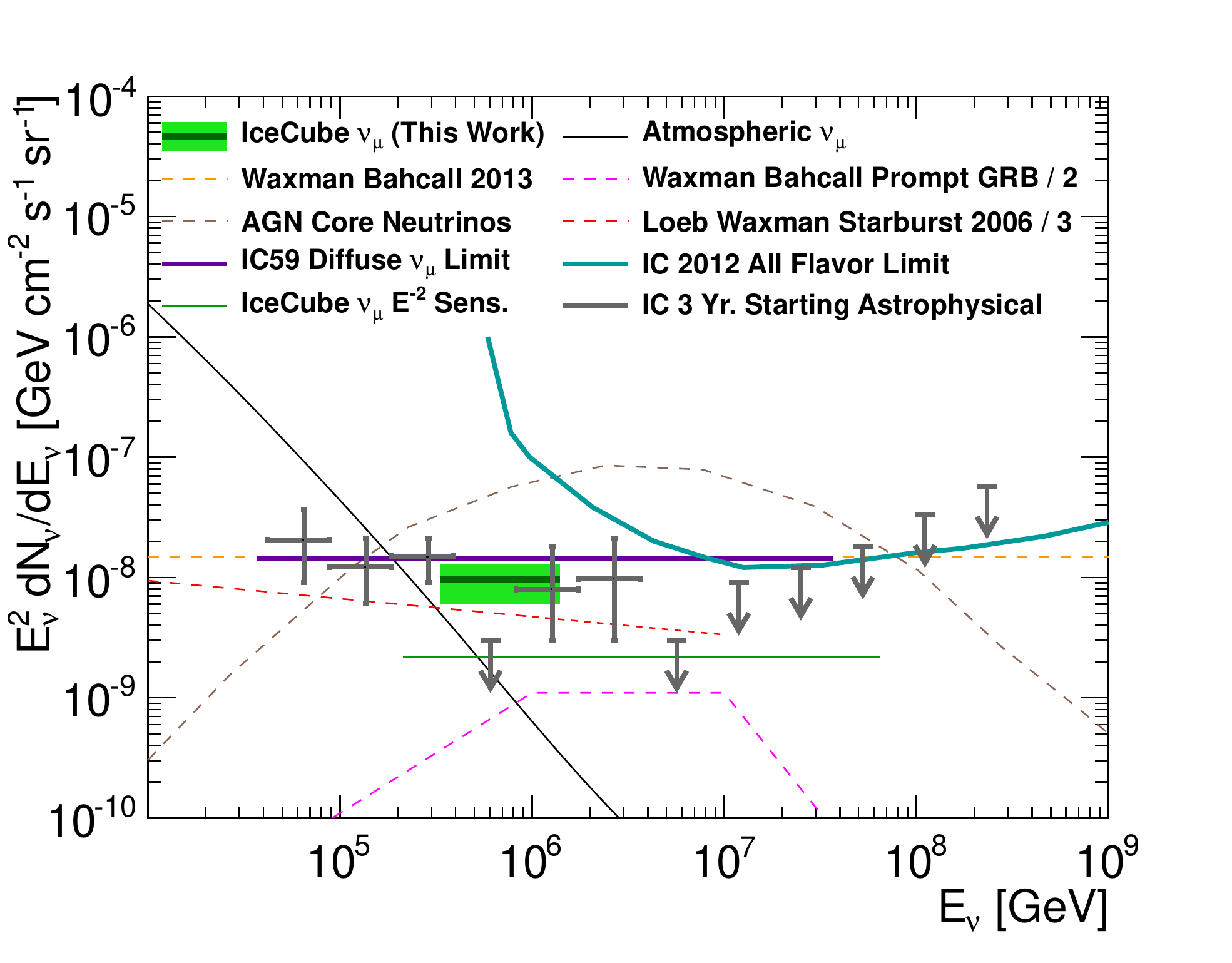}\label{fig:ICdetection}}
}
\caption{(a) The upper bound imposed by UHECR observations on the extra-Galactic (all flavor) high energy neutrino intensity (lower-curve: no evolution of the energy production rate, upper curve: assuming evolution following the star formation rate), eq.~\ref{eq:WB}, compared with the atmospheric neutrino background. The curve labelled "GZK" shows the neutrino intensity expected from UHECR proton interactions with micro-wave background photons\cite{gzk_nu}. The dash-dotted blue line shows the muon neutrino intensity that would produce one neutrino event per year in a detector with an effective mass of 1~Gton.
(b) The extra-terrestrial neutrino signal (single flavor, assuming a $\nu_e:\nu_\mu:\nu_\tau=1:1:1$ flavor ratio) detected by IceCube in the energy range of $\sim50$~TeV to $\sim2$~PeV (points with error bars- "starting events" analysis\cite{IC14PhRvL_3yr_detection}; green shaded area- muon analysis\cite{IC15_nu_mu_detection}; adopted from ref.~\refcite{IC15_nu_mu_detection}). Within the current relatively large uncertainties, the detected extra-terrestrial signal coincides with the WB bound (eq.~\ref{eq:WB}). Theroetical model predictions of the emission from particular astrophysical sources (starbursts\cite{LW06_StarBurst_nus}, GRBs\cite{WnB97}, AGNs\cite{Stecker05_AGN_core}) are also shown.
}
\label{fig:Bound_Detection} 
\end{figure}

Fig.~\ref{fig:WBbound} illustrates that a giga-ton neutrino telescope is required to detect the expected flux in the energy range of $\sim1$~TeV to $\sim1$~PeV, and that a much larger effective mass is required at higher energy. For a proton dominate UHECR flux, the WB bound is expected to be saturated at $\sim10^{10}$~GeV, since most of the UHECR protons that have been produced over a Hubble time would have lost all their energy to pion production in interactions with the CMB photons\cite{gzk_nu}.

\section{The origin of IceCube's neutrinos}\label{sec:IC_nu_origin}

\subsection{The extra-terrestrial component of the neutrino flux}\label{sec:IC_detection}

The IceCube collaboration reported\cite{IC14PhRvL_3yr_detection} a detection of 37 neutrinos in the energy range of $\sim50$~TeV to $\sim2$~PeV, which constitutes a $5.7\sigma$ excess above the expected atmospheric neutrino and muon backgrounds (the uncertain contribution of atmospheric neutrinos from charmed meson decay is constrained by the
angular distribution of the detected events). The excess neutrino spectrum is consistent with a "flat", $E_\nu^{2}dn_\nu/dE_\nu \propto E_\nu^{0}$, spectrum, its angular distribution is consistent with isotropy, and its flavor content is consistent with $\nu_e:\nu_\mu:\nu_\tau=1:1:1$\cite{IC15PhRvL_flavor}. Assuming a flat spectrum, the best fit normalization of the intensity is $E_\nu^2\Phi_\nu=(2.85\pm0.9)\times10^{-8}{\rm GeV/cm^2 s\, sr}$. Assuming that this intensity extends to high energy, 3 events should have been detected on average above 2~PeV. The absence of such events suggests a suppression of the flux above 2~PeV, or a softer than $E_\nu^{2}dn_\nu/dE_\nu \propto E_\nu^{0}$ spectrum. Fitting a power law excess, $dn_\nu/dE_\nu \propto E_\nu^{-\alpha}$ extending beyond 2~PeV , to the data yields $\alpha=2.3\pm0.3$ (90\% confidence).

The results of ref.~\refcite{IC14PhRvL_3yr_detection} were obtained by an analysis limited to events for which the neutrino interaction occurs within the detector ("starting events"). A recent analysis\cite{IC15_nu_mu_detection} searching for high energy neutrino induced muon events, not limited to neutrino interactions within the detector, revealed a muon-neutrino flux which constitutes a $3.7\sigma$ excess above the expected atmospheric neutrino and muon backgrounds in the energy range of $\sim300$~TeV to $\sim1$~PeV, with flux and spectrum consistent with those of the "starting events" analysis.
The results of the "starting events" and muon analyses described above are shown graphically in fig.~\ref{fig:ICdetection}.

Extending the analysis to lower energy, $\sim10$~TeV, where an astrophysical signal would be strongly dominated by the atmospheric flux, an excess of events is found at $\sim30$~TeV\cite{IC15_10TeV} above an extension of the flat, $E_\nu^{2}dn/dE_\nu \propto E_\nu^{0}$, $>50$~TeV spectrum. This may indicate a new low energy component or a steeper ($\alpha\approx2.5$) spectrum across the entire observed range. Robust conclusions cannot yet be drawn, since the significance of the excess is not high, $2\sigma$\cite{IC15_combined_likelyhood}, and it is sensitive to the choice of energy bins (e.g. fig.~12 of ref.~[\refcite{IC15_10TeV}]). The possible low energy excess does not affect the analysis presented here, which is focused on the higher energy, $>50$~TeV, neutrinos.

A $\nu_e:\nu_\mu:\nu_\tau=1:1:1$ flavor ratio is consistent with that expected for pion decay in cosmologically distant sources, for which oscillations modify the original $1:2:0$ ratio to a $1:1:1$ ratio\cite{Learned95}. However, with the current limited statistics, the data are consistent with any initial (i.e. at the source) flavor ratio\cite{IC15PhRvL_flavor}.

\subsection{UHECR sources in "cosmic calorimeters"}\label{sec:calorimeters}

The extra-terrestrial neutrino flux is unlikely to be dominated by (yet unknown) Galactic sources, which, unlike the observed signal, are expected to be strongly concentrated along the Galactic disk. This, and the coincidence with the WB bound, suggests an extra-Galactic origin. IceCube's neutrinos are therefore most likely emitted by the decay of charged pions produced in interactions of high energy CRs with ambient (low energy) photons or protons in extra-Galactic sources.

Let us consider first the case where the parent CRs are protons. The fraction of the parent proton energy carried by each neutrino is approximately $1/20$ (for production by interaction with either photons or protons). Since the neutrino flux coincides with the bound given in eq.~\ref{eq:WB} over the $\sim50$~TeV to $\sim2$~PeV energy range, a lower limit of $E_p^2d\dot{n}_p/dE_p\ge Q_{\rm UHE}$ is implied on the local, $z=0$, proton production rate in the energy range of $\approx 1$~PeV to $\approx50$~PeV. Two distinct scenarios may be considered. The observed neutrinos may be produced by sources accelerating protons at a rate $E_p^2d\dot{n}_p/dE_p\sim Q_{\rm UHE}$, provided that protons of energy $E_p<50$~PeV lose most of their energy to pion production either within the sources or at the sources' environment. Alternatively, the observed neutrinos may be produced by sources accelerating protons at a rate $(E_p^2d\dot{n}_p/dE_p)\gg Q_{\rm UHE}$, provided that protons lose only a small fraction, $f(E)\ll1$, of their energy to pion production. In the latter case, the small (and likely energy dependent) energy loss fraction $f(E)$ should compensate the large energy production rate, $(E_p^2d\dot{n}_p/dE_p)/Q_{\rm UHE}\gg1$, to reproduce the observed flux and spectrum over two decades of $\nu$ energy, and the coincidence of the observed neutrino flux and spectrum with the WB bound would be an accident.

The simpler explanation, which we consider to be more likely, is that both the neutrinos and the UHECRs are produced by the same population of cosmological sources, producing protons with a flat spectrum, $d\log\dot{n}/d\log E\approx -2$ as observed in a wide range of astrophysics accelerators and as expected theoretically for electromagnetic acceleration in collisionless shocks (see \S~\ref{sec:UHECRs}), over the [1~PeV,$10^{11}$~GeV] energy range, and residing in "calorimetric" environments, in which protons of energy $E_p< 50$~PeV lose much of their energy to pion production.

Let us consider next the case where the parent CRs are heavy nuclei of atomic mass $A$. The fraction of the parent nucleus energy carried by each neutrino emitted by the decay of charged pions produced in inelastic collisions with ambient protons is approximately $1/20A$, while interactions with ambient photons may lead to photo-disintegration of the nuclei, roughly preserving $E/A$. In this case therefore, a lower limit of $E_A^2d\dot{n}_A/dE_A\ge Q_{\rm UHE}$ is implied on the local, $z=0$, production rate of nuclei in the energy range of $\approx 1A$~PeV to $\approx50A$~PeV. As mentioned in \S~\ref{sec:UHECRs}, if the UHECR flux is dominated by heavy nuclei, the energy production rate required to account for the observed UHECR flux would differ from that given by eq.~\ref{eq:uhecr} by a factor of a few at $10^{10.5}$~GeV. Thus, the observed neutrino flux and spectrum and the CR flux at $10^{10.5}$~eV could be explained by a single population of sources producing heavy nuclei with a flat spectrum, $d\log\dot{n}/d\log E\approx -2$ over the [1$A$~PeV,$10^{11}$~GeV] energy range, and residing in "calorimetric" environments, in which nuclei of energy $E/A\approx E/2Z< 50$~PeV lose much of their energy to pion production.

Unlike the case of protons, if the UHECR flux is dominated by heavy nuclei, the $>10^{10}$~GeV spectrum cannot be simply explained by an extension of a flat, $d\log\dot{n}/d\log E\approx -2$, spectrum to high energy (see \S~\ref{sec:UHECRs}). We consider this an additional evidence
supporting a proton dominated UHECR flux.

For all production channels, $p(A)-p(\gamma)$, similar flavor content ($1:1:1$) and particle/anti-particle content ($I_\nu=I_{\bar{\nu}}$) are expected, except for $p\gamma$ interactions dominated by the $\Delta$ resonance (which may be obtained in environments with soft, $d\log n_\gamma/d\log E_\gamma<-1$, photon spectra), where an excess of particles over anti-particles is expected.

Finally, we note that the absence of neutrino detection above a few PeV, which suggests a suppression of the neutrino flux above this energy, may be due to efficient escape of $E/Z>100$~PeV CRs from the environments in which they produce the pions (as was predicted to be the case for sources residing in starburst galaxies\cite{LW06_StarBurst_nus}), and need not imply a cutoff in the CR production spectrum (see \S~\ref{sec:starbursts}).

\subsection{Star forming galaxies}\label{sec:sfr_galaxies}

\subsubsection{Starburst calorimeters}\label{sec:starbursts}

Starburst galaxies have been predicted\cite{LW06_StarBurst_nus} to act as cosmic-calorimeters, producing an extra-Galactic neutrino background comparable to the WB bound at energies $E<0.5$~PeV, and possibly extending to higher energy (see fig.\ref{fig:ICdetection}). Starbursts may be defined as galaxies with specific star formation rate (sSFR), i.e. SFR per galaxy stellar mass, which is much higher than the average sSFR of galaxies at a similar redshift\cite{Sarget12SB_fraction}. Starbursts in the local universe are characterized by disks of typical radii $\ell$ of several hundred parsecs, with column densities $\Sigma_g>0.1{\rm g/cm}^2$\cite{Voelk89Calorimetry,Condon91SB_Properties,Kennicutt98} and magnetic fields $B\sim1$~mG ($B\propto\Sigma_g$)\cite{TQWetal06SB_B}, which are much larger than those of "normal" spiral galaxies ($\Sigma_g\approx0.003{\rm g/cm}^2$, $B\sim5\,\mu$G in the Milky way). The large disk densities imply that the energy loss time of CR protons, due to inelastic $pp$ collisions, is much shorter in starburst galaxies than in normal spirals, and the enhanced magnetic field implies that the confinement time of the protons is expected to be larger than in normal spirals. This in turn implies that, unlike normal spirals, starburst galaxies may act as proton calorimeters.

Starburst galaxies have long been argued\cite{Voelk89Calorimetry} to act as calorimeters for few~GeV protons, based on the FIR-radio correlation. The recent detection of GeV and TeV emission from the nearby starburst galaxies M82 and NGC253 indicate that these galaxies are calorimetric for protons of energy exceeding 10~TeV\cite{Lacki11GeV_SB_detection}. The theoretical arguments given in ref.~\refcite{LW06_StarBurst_nus}, and reproduced in a more general way below, suggest that calorimetry holds to energies exceeding 10~PeV.

Protons will lose all their energy to pion production provided that the energy loss time is shorter than both the starburst lifetime and the magnetic confinement time within the starburst gas. In the energy range of interest, the inelastic nuclear collision cross section is $\sigma_{\rm pp}\approx50$~mb, with inelasticity of $\approx0.5$. The energy loss time, $\tau_{\rm loss}\approx (0.5 n\sigma_{\rm pp}c)^{-1}$ where $n$ is the interstellar nucleon density, would be shorter than the starburst lifetime, which is at least the dynamical time given by the ratio of $\ell$ to the characteristic gas velocities $v$, $\sim(2\ell/v)$, as long as
\begin{equation}\label{eq:Sigma_c}
\Sigma_{\rm gas}\gtrsim \Sigma_{\rm crit}\equiv {m_p v \over \sigma_{\rm
pp}c}= 0.03 (v/300~{\rm km~s^{-1}})~{\rm g~cm^{-2}}.
\end{equation}
For characteristic gas velocities, $v=$ few hundred km/s, the critical surface density, $\Sigma_{\rm crit}$, is comparable to the minimum $\Sigma_{\rm gas}$ of known starburst galaxies\cite{Voelk89Calorimetry,TQWetal06SB_B}.

The ratio of confinement time and loss time is less straightforward to estimate, since magnetic confinement of CRs is not well understood. In the Milky Way, the total gas column density traversed by CRs of energy $E/Z\le1$~TeV before they escape the Galaxy is $\Sigma_{\rm conf, MW}\approx9(E/10Z{\rm GeV})^{-0.4}{\rm g~cm^{-2}}$\cite{Blandford87,BKW13AMS02}. If the propagation of CRs in starburst galaxies is similar to that in our Galaxy, we may expect $\Sigma_{\rm conf,SB}(E/Z)=(n_{\rm SB}/n_{\rm MW})\Sigma_{\rm conf,MW}(B_{\rm MW}E/B_{\rm SB}Z)$, since the propagation of the CRs in the magnetic field is determined by $E/ZB$. For $n_{\rm SB}/n_{\rm MW}=B_{\rm SB}/B_{\rm MW}=100$, the fraction of proton energy lost to pion production before escape is $f_{\pi,\rm SB} \approx 1(E/1{\rm PeV})^{-0.4}$.
Since the neutrino flux is expected to be dominated by starbursts at $z\gtrsim 1$, for which the typical surface density should be even higher than in local starbursts, it is reasonable to assume that most of the energy injected into starburst galaxies in $E\lesssim10$~PeV protons is converted to pions.

\subsubsection{The fraction of CR production occurring in calorimetric galaxies}\label{sec:calorimetric_fraction}

The energy loss of $E\lesssim10$~PeV protons in starburst galaxies would produce a neutrino flux and intensity similar to the WB bound at $E\lesssim1$~PeV, provided that the sources of UHECRs produce most of their energy output in starburst environments. This would be the case if, as assumed in ref.~\refcite{LW06_StarBurst_nus}, (i) the rate of CR production is proportional to the SFR and (ii) most (or a significant fraction) of the stars in the universe have been produced in starburst episodes.

The leading UHECR accelerator candidates are $\gamma$-ray bursts (GRBs) and transient accretion events onto massive black holes residing at the centers of galaxies\cite{Lemoine13,W11_frontiers_review}. The assumption that the CR production rate is proportional to the SFR is natural for sources like GRBs, which are related to the deaths of massive stars, and may not be inconsistent with models based on activity around massive central black holes. On the other hand, the assumption that a significant fraction of the stars in the universe have been produced in starburst episodes (as suggested e.g. by ref.~\refcite{Juneau05SB_fraction}) is not necessarily valid.

Recent observations indicate that for $z<2$ the sSFR is narrowly distributed around a $z$-dependent average, which increases by a factor of $\sim30$ from $z=0$ to $z=2$\cite{Sarget12SB_fraction}. Outliers with high sSFR, which may be categorized as starbursts, are found to contribute only $\sim10$\% of the total SFR\cite{Sarget12SB_fraction}. The interpretation of these results is still debated. In particular, starburst activity has been commonly argued to be triggered by galaxy mergers, and whether or not the narrow distribution of the sSFR rules out major mergers as the cause for the rapid increase of the average sSFR with redshift is still debated\cite{Daddi10Massive_disk_z1.5,Puech14major_mergers}. In other words, if starburst activity is driven by mergers and major mergers are responsible for the increase in the sSFR, all galaxies at higher redshift may be classified as starbursts.

While the definition of a starburst is debateable and the fraction of star-formation occurring in starburst episodes is still uncertain, it should be noted that the typcial galaxies at high $z$, which are rapidly forming stars, may well be calorimetric. CO observations of 6 $z=1.5$ galaxies with 'normal' (i.e. close to the average) sSFR show that they are characterized by rapid SFR, $\sim 100M_\odot/{\rm yr}$, and high column density massive molecular disks, $\Sigma_g\sim0.1{\rm g/cm^2}$\cite{Daddi10Massive_disk_z1.5}. In the local universe, galaxies with such column density and SFR are calorimetric(see eq.~\ref{eq:Sigma_c} and ref.~\refcite{Lacki11GeV_SB_detection}). While the disk structure of high $z$ galaxies may be different, and the determination of the high $z$ molecular gas content is uncertain, these results support the hypothesis that a large fraction of the SFR occurred in calorimetric environments.

\subsection{A lower limit to the density of steady sources}\label{sec:source_density}

The non detection by IceCube of point sources producing multiple neutrino events, combined with the measured "diffuse" neutrino intensity, sets a lower limit to the density of the sources producing the neutrinos, and an upper limit to their neutrino luminosity. We give below a simple order of magnitude estimate of these limits.

Consider a population of "standard candle" sources, with density $n_s(z)$ and 0.1 to 1~PeV muon neutrino luminosity $L_{\nu_\mu}$. We consider only neutrino induced muon events, for which the arrival direction may be determined with good accuracy (better than 1~deg). The coincidence of the neutrino intensity with the WB bound implies, using eq.~\ref{eq:WB} and a measured $1:1:1$ flavor ratio,
\begin{equation}\label{eq:specific_nu_L}
  n_0L_{\nu_\mu}\approx 2\times10^{43}\left(\frac{\xi_z}{3}\right)^{-1}{\rm erg/Mpc^3yr},
\end{equation}
where $n_0\equiv n_s(z=0)$. The flux of neutrinos of energy $E_\nu$ required for the detection of more than one neutrino induced muon event is approximately given by $f_{\rm m}=E_\nu/ATP_{\mu\nu}$, where $A$ is the detector's effective area, $T$ is the integration time, and the probability that a 1~TeV to 1~PeV neutrino with a propagation track crossing the detector would produce a muon going through the detector is $P_{\mu\nu}\approx10^{-6}(E_\nu/1{\rm TeV})$\cite{Gaisser95_PhysRep}. This yields
$f_m\approx2\times10^{-12}(AT/3{\rm km^2yr})^{-1}{\rm erg/cm^2s}$.
The limit inferred below on $n_0$ implies that the distance $d_m$ below which the flux exceeds $f_m$, $d_m=(L_{\nu_\mu}/4\pi f_m)^{1/2}$, is  $\ll c/H_0$. Thus, the number of sources producing multiple upward moving (and hence neutrino induced) muon events is approximately
\begin{equation}\label{eq:Nm}
  N_m\approx\frac{2\pi}{3}n_0d_m^3\approx 1\left(\frac{\xi_z}{3}\right)^{-3/2}\left(\frac{n_0}{10^{-7}{\rm Mpc}^{-3}}\right)^{-1/2}\left(\frac{A}{1\rm km^2}\frac{T}{3\rm yr}\right)^{3/2}.
\end{equation}

The requirement $N_m<1$ sets a lower limit to the density of steady sources, $n_0>10^{-7}{\rm Mpc}^{-3}$ (implying $L_{\nu_\mu}<10^{43}{\rm erg/s}$ and $d_m<200$~Mpc), consistent with that of the detailed analysis of ref.~\refcite{AH14source_density}. It rules out rare candidate sources, like bright $L\sim10^{47}{\rm erg/s}$ AGN with $n_0\sim10^{-9}{\rm Mpc}^{-3}$, and is consistent with the density of starburst galaxies, $n_0\sim10^{-5}{\rm Mpc}^{-3}$.

\section{Summary and discussion}\label{sec:discussion}

\subsection{The calorimetric star-forming galaxies model and its uncertainties}\label{sec:calorimetic_summ}

IceCube's measurements of extra-terrestrial neutrinos are consistent with a model in which both the neutrinos and the observed UHECRs are produced by the same population of cosmological sources, producing CR protons at a similar rate, $E^2d\dot{n}/dE\propto E^{0}$, over the [1~PeV,$10^{11}$~GeV] energy range, and residing in "calorimetric" environments, in which $E<50$~PeV protons lose much of their energy to pion production (\S~\ref{sec:calorimeters}). This model is a natural explanation of IceCube's results since it relies on a known (although not yet identified) population of sources- the UHECR sources, and since it does not depend on ad-hoc choices or fine tuning of model parameters: the neutrino flux normalization is determined by the observed UHECR flux, and the neutrino spectrum is consistent with that implied by the measured UHECR spectrum (\S~\ref{sec:UHECRs}). Moreover, a "flat" $d\log\dot{n}/d\log E\approx -2$ generation spectrum is observed in a wide range of astrophysical environments\cite{Blandford87,Axford94,Waxman_rel-plasma_rev06} and is a robust prediction of the most widely accepted and best understood (although not yet from first principles) model for particle acceleration in astrophysical objects- Fermi acceleration in collisionless shocks\cite{Blandford87,Bednarz98RelFermi,Keshet05RelFermi}.

A wide variety of (non calorimetric) models for the neutrino origin were proposed following IceCube's discovery (including dark matter decay, active galactic nuclei of various types, see fig.~\ref{fig:ICdetection}, and galaxy clusters, see refs.~\refcite{Meszaros14Rev,Murase15review} for reviews). These models generally rely on ad-hoc choices of model parameters, which cannot be derived theoretically or determined observationally, in order to reproduce the observed flux and spectrum of the neutrinos. Their predictive power is thus limited.

Radio to TeV observations of local starburst galaxies imply that they are calorimetric for protons of energy exceeding 10~TeV\cite{Voelk89Calorimetry,Lacki11GeV_SB_detection}, and theoretical arguments suggest that calorimetry holds to energies exceeding 10~PeV (\S~\ref{sec:starbursts}). Starburst galaxies have thus been predicted\cite{LW06_StarBurst_nus} (fig.~\ref{fig:ICdetection}) to produce an extra-Galactic neutrino background comparable to the WB bound at energies $E<0.5$~PeV, and possibly extending to higher energy, provided that (i) the rate of CR production is proportional to the SFR and (ii) most (or a significant fraction) of the stars in the universe have been produced in starburst environments. The validity of the first assumption is natural if UHECRs are produced by GRBs (and may hold also for models based on activity around massive central black holes). It is further supported by the observation that at low CR energy, $\sim10$~GeV, the ratio of CR production rate to SFR is similar, $\sim10^{47}{\rm erg}$ for 1~$M_\odot$ of star formation, in the Galaxy and in starburst galaxies, while their SFRs differ by orders of magnitude\cite{Katz13SingleSource}.

The production rate per unit volume of CR protons at $\sim10$~GeV, $E^2d\dot{n}/dE|_{10\,\rm GeV}$, is only $\sim10$ times the rate at UHE, suggesting that the same sources are responsible for the production of CRs of all energies\cite{Katz13SingleSource}, from 1~GeV to $10^{11}$~GeV (implying $d\log\dot{n}/d\log E\simeq-2.1$). Alternatively, there may exist a population of sources, associated with star formation and embedded in starbursts (like supernovae), producing CRs at a rate and spectrum similar to that of the UHE sources but reaching only lower energy ($\ll10^{10}$~GeV), and thus making a contribution to the lower energy CR flux and possibly to the 1~PeV neutrino flux\cite{He13SB_HNe,Liu14SB_HNe,Senno15SB_HNe}, which is similar to that of the UHECR sources. In this scenario, the similarity of the energy production rates of the lower $E$ and UHE sources requires an explanation\footnote{
The ad-hoc explanation given in the next to last parag. of ref.~\refcite{LW06_StarBurst_nus} is incorrect. It follows erroneous statements that are inconsistent with the rest of that article (That the local 1.4~GHz luminosity is dominated by starbursts, and that UHECRs do not escape starbursts)}.

The main uncertainty remaining is the fraction of stars formed in calorimetric environments. The observed neutrino intensity is dominated by sources at redshift 1--2. The 'average' galaxies at these redshifts produce stars at a much higher rate than local galaxies\cite{Sarget12SB_fraction}, and observations indicate that they are characterized by high column density molecular disks, $\Sigma_g\sim0.1{\rm g/cm^2}$\cite{Daddi10Massive_disk_z1.5}. While this provides an indication that a significant fraction of the star formation occurred in calorimetric environments (\S~\ref{sec:sfr_galaxies}), the structure of galaxy disks at this redshift range is poorly constrained.

The production of neutrinos by pion decay is accompanied by the production of high energy photos, which initiate electromagnetic cascades via interaction with infra-red background photons, leading to a background of $\lesssim0.1$~TeV photons. In calorimetric models with a power-law proton spectrum, pion production by $\sim1$~TeV protons contributes directly to the $\lesssim0.1$~TeV photon background. The resulting background intensity is expected to contribute a significant fraction of the observed 0.1~TeV background\cite{TQW07SB_gamma}, limiting the proton spectra to $d\log\dot{n}/d\log E>-2.2$\cite{Murase13100GeVconst,Chang14SB_gbgnd,Tamborra14SB_gbgnd}.

\subsection{Open questions and the way towards their resolution}\label{sec:Qprospects}

Due to the limited statistics and flavor discrimination power, the current uncertainties in the determination of the neutrino spectrum, angular distribution and flavor content are large (\S~\ref{sec:IC_detection}). A significant reduction of uncertainties requires a significant (order of magnitude) expansion of the effective mass of the detector at $\gtrsim100$~TeV. Such expansion is necessary for example for a study of the hints for spectral breaks at low, $<30$~TeV, and high, $>2$~PeV energy, which are currently detected with low statistical significance. Reduced uncertainties will provide much more stringent constraints on predictive models, such as the calorimetric model.

An identification of the sources by angular correlation with a catalog of nearby sources is unlikely, since the fraction of events originating from $z<0.1$ sources is $\approx1/20$, implying that the fraction of well localized neutrino induced muon events from $z<0.1$ sources over $2\pi$~sr is $\sim1/200$.

A detection of neutrino emission from few nearby bright starburst galaxies would constitute a major evidence in support of the calorimetric star-forming galaxy model. Since the local density of starbursts is $n_0\sim10^{-5}{\rm Mpc^{-3}}$, a $\sim10$-fold increase in the effective area of the detector at $\sim100$~TeV is required in order to enable the detection of a few nearby sources (see eq.~\ref{eq:Nm} and refs.~\refcite{AH14source_density,Anchordoqui14SB_nu_detection}).

An identification of the neutrino sources is likely to identify the calorimeters within which the CR accelerators reside, but not the accelerators themselves. The neutrino flux that is produced within the accelerators is expected to be significantly lower than the total neutrino flux produced in the calorimeters surrounding them. For example, if UHECRs are protons produced in GRBs, the neutrino flux expected to be produced within the GRB accelerators is $\approx10$\% ($\approx1$\%) of the WB flux at 1~PeV (0.1~PeV)\cite{WnB97} (fig.~\ref{fig:ICdetection}). Such a low flux would imply that had the accelerators been steady sources, even a tenfold increase in detector mass would have been unlikely to enable their identification. Luckily, the UHECR accelerators must be transient: The absence of (steady) sources with power output exceeding the minimum required for proton acceleration to $10^{11}$~GeV, $L>10^{46}{\rm erg/s}$, within the $\sim100$~Mpc propagation distance of such high energy protons, implies that the sources must be extremely bright transients\cite{W11_frontiers_review,Lemoine13}. Their identification may be possible by an association of a neutrino with an electromagnetic signal accompanying the transient event. In order to open the possibility for such associations, a tenfold increase in detector mass and a wide field electromagnetic transient monitoring program are required.

A large sample of high energy neutrinos may enable one to study both neutrino properties and possibly deviations from the standard model of particle physics (see ref.~\refcite{Murase15review} for a recent summary of such possibilities). The identification of the sources is important for such studies, in order to discriminate between spectral/flavor features originating from astrophysical and particle physics model effects (e.g. ref.~\refcite{Kashti08}).

Finally, while we have argued (\S~\ref{sec:UHECRs}, \S~\ref{sec:calorimeters}) that a proton dominated composition is more likely, a heavy nuceli dominated composition cannot be excluded. A detection of (or stringent upper limit on) the GZK neutrino flux predicted for a proton dominated composition (\S~\ref{sec:WBder}, fig.~\ref{fig:WBbound}) will provide a clear confirmation of (or will rule out) a proton dominated composition. The recent analysis of Auger\cite{Auger15GZK_nu_limit} sets an upper limit which is close to the WB bound at $\sim10^{9.5}$~GeV, and the expected flux may be detectable by future radio experiments\cite{Gorham13Ballon,Allison14ARA,Barwick15AIANNA}.

\bibliographystyle{ws-rv-van}

\begin{thebibliography}{58}
\providecommand{\natexlab}[1]{#1}
\providecommand{\url}[1]{\texttt{#1}}
\expandafter\ifx\csname urlstyle\endcsname\relax
  \providecommand{\doi}[1]{doi: #1}\else
  \providecommand{\doi}{doi: \begingroup \urlstyle{rm}\Url}\fi

\bibitem{Helder12_SNR_CR}
E.~A. {Helder}, et al. \emph{Space Sci. Rev.} {\bf 173}, \penalty0 369--431 (Nov., 2012).

\bibitem{Lemoine13}
M.~{Lemoine},
  \emph{Journal of Physics Conference Series}. {\bf 409}\penalty0 (1):\penalty0
  012007 (Feb., 2013).

\bibitem{W11_frontiers_review}
E.~{Waxman}, in Astronomy at the Frontiers of Science, ed. J.-P. Lasota, Springer (Aug., 2011) (arXiv:1101.1155).

\bibitem{Gaisser95_PhysRep}
T.~K. {Gaisser}, F.~{Halzen}, and T.~{Stanev}, \emph{Phys. Rep.} {\bf 258}, \penalty0 173--236 (July,
  1995).

\bibitem{WBbound1}
E.~{Waxman} and J.~{Bahcall}, \emph{\prd}. {\bf 59}\penalty0 (2), \penalty0
  023002--+ (Jan., 1999).

\bibitem{WBbound2}
J.~{Bahcall} and E.~{Waxman}, \emph{\prd}. {\bf 64}\penalty0 (2), \penalty0 023002--+
  (July, 2001).

\bibitem{IC14PhRvL_3yr_detection}
M.~G. {Aartsen},
  et~al., \emph{PRL}. {\bf113}\penalty0 (10):\penalty0
  101101 (Sept., 2014).

\bibitem{IC15_nu_mu_detection}
{IceCube Collaboration}, M.~G. {Aartsen}, et~al., arXiv:1507.04005 (July. 2015).

\bibitem{Bird93_FE_comp}
D.~J. {Bird}, et~al., \emph{PRL}. {\bf 71},
  \penalty0 3401--3404 (Nov., 1993).

\bibitem{HiRes_Composition_05}
R.~U. {Abbasi et al. (HiRes Collaboration)},
  \emph{ApJ}. {\bf 622}, \penalty0 910--926 (Apr., 2005).

\bibitem{TA_Comp_2015}
R.~U. {Abbasi}, et~al., \emph{Astroparticle Physics}. {\bf 64},
  \penalty0 49--62 (Apr., 2015).

\bibitem{Auger_comp_2014}
A.~{Aab}, et~al., \emph{\prd}. {\bf 90}\penalty0
  (12):\penalty0 122006 (Dec., 2014).

\bibitem{LW09_Aniso_Comp}
M.~{Lemoine} and E.~{Waxman}, \emph{JCAP}. {\bf 11}, \penalty0 9--+ (Nov., 2009).

\bibitem{W95_QUHECR}
E.~{Waxman},
  \emph{ApJ}. {\bf 452}, \penalty0 L1+ (Oct., 1995).

\bibitem{BW03_QUHECR}
J.~N. {Bahcall} and E.~{Waxman},
  \emph{Physics Letters B}. {\bf 556}, \penalty0 1--6 (Mar., 2003).

\bibitem{Katz_UHECR_09}
B.~{Katz}, R.~{Budnik}, and E.~{Waxman}, \emph{JCAP}. {\bf 3}, \penalty0 20--+ (Mar.,
  2009).

\bibitem{Greisen}
K.~{Greisen}, \emph{PRL}. {\bf 16}, \penalty0 748--750 (Apr., 1966).

\bibitem{ZK}
G.~T. {Zatsepin} and V.~A. {Kuz'min}, \emph{JETP Letters}.
  {\bf 4}, \penalty0 78--+ (Aug., 1966).

\bibitem{Blandford87}
R.~{Blandford} and D.~{Eichler}, \emph{Phys. Rep.} {\bf 154},
  \penalty0 1--75 (Oct., 1987).

\bibitem{Axford94}
W.~I. {Axford}, \emph{ApJ
  Supp.} {\bf 90}, \penalty0 937--944 (Feb., 1994).

\bibitem{Waxman_rel-plasma_rev06}
E.~{Waxman}, \emph{Plasma Physics
  and Controlled Fusion}. {\bf 48}, \penalty0 B137--B151 (Dec., 2006).

\bibitem{Bednarz98RelFermi}
J.~{Bednarz} and M.~{Ostrowski}, \emph{PRL}. {\bf 80},
  \penalty0 3911--3914 (May, 1998).

\bibitem{Keshet05RelFermi}
U.~{Keshet} and E.~{Waxman}, \emph{PRL}.
  {\bf 94}\penalty0 (11):\penalty0 111102 (Mar., 2005).

\bibitem{Allard12}
D.~{Allard},
  \emph{Astroparticle Physics}. {\bf 39}, \penalty0 33--43 (Dec., 2012).

\bibitem{gzk_nu}
V.~S. {Beresinsky} and G.~T. {Zatsepin}, \emph{Physics Letters B}. {\bf 28}, \penalty0 423--424 (Jan.,
  1969).

\bibitem{LW06_StarBurst_nus}
A.~{Loeb} and E.~{Waxman}, \emph{\jcap}. {\bf 5}:\penalty0 003 (May, 2006).

\bibitem{WnB97}
E.~{Waxman} and J.~{Bahcall}, \emph{PRL}. {\bf 78},
  \penalty0 2292--2295 (Mar., 1997).

\bibitem{Stecker05_AGN_core}
F.~W. {Stecker}, \emph{\prd}. {\bf 72}\penalty0 (10):\penalty0 107301 (Nov., 2005).

\bibitem{IC15PhRvL_flavor}
M.~G. {Aartsen}, et~al.,
  \emph{PRL}. {\bf 114}\penalty0 (17):\penalty0 171102 (May,
  2015).

\bibitem{IC15_10TeV}
M.~G. {Aartsen}, et~al., \emph{\prd}. {\bf 91}\penalty0 (2):\penalty0 022001 (Jan., 2015).

\bibitem{IC15_combined_likelyhood}
M.~G. {Aartsen}, et~al., \emph{\apj}.
  {\bf 809}:\penalty0 98 (Aug., 2015).

\bibitem{Learned95}
J.~G. {Learned} and S.~{Pakvasa}, \emph{Astroparticle Physics}. {\bf 3}, \penalty0 267--274 (May,
  1995).

\bibitem{Sarget12SB_fraction}
M.~T. {Sargent}, et~al., \emph{\apjl}. {\bf 747}:\penalty0 L31 (Mar., 2012).

\bibitem{Voelk89Calorimetry}
H.~J. {Voelk}, \emph{\aap}. {\bf 218}, \penalty0
  67--70 (July, 1989).

\bibitem{Condon91SB_Properties}
J.~J. {Condon}, et~al., \emph{\apj}. {\bf 378},
  \penalty0 65--76 (Sept., 1991).

\bibitem{Kennicutt98}
R.~C. {Kennicutt}, Jr.,
  \emph{\apj}. {\bf 498}, \penalty0 541--552 (May, 1998).

\bibitem{TQWetal06SB_B}
T.~A. {Thompson}, et~al., \emph{\apj}. {\bf 645}, \penalty0 186--198 (July, 2006).

\bibitem{Lacki11GeV_SB_detection}
B.~C. {Lacki}, et~al.,
  \emph{\apj}. {\bf 734}:\penalty0 107 (June, 2011).

\bibitem{BKW13AMS02}
K.~{Blum}, B.~{Katz}, and E.~{Waxman}, \emph{PRL}.
  {\bf 111}\penalty0 (11):\penalty0 211101 (Nov., 2013).

\bibitem{Juneau05SB_fraction}
S.~{Juneau}, et~al., \emph{\apjl}.
  {\bf 619}, \penalty0 L135--L138 (Feb., 2005).

\bibitem{Daddi10Massive_disk_z1.5}
E.~{Daddi}, et~al., \emph{\apj}. {\bf 713}, \penalty0
  686--707 (Apr., 2010).

\bibitem{Puech14major_mergers}
M.~{Puech}, et~al., \emph{\mnras}. {\bf 443}, \penalty0 L49--L53
  (Sept., 2014).

\bibitem{AH14source_density}
M.~{Ahlers} and F.~{Halzen}, et~al., \emph{\prd}. {\bf 90}\penalty0 (4):\penalty0
  043005 (Aug., 2014).

\bibitem{Meszaros14Rev}
P.~{M{\'e}sz{\'a}ros}, \emph{Nuclear Physics B
  Proc. Supp.} {\bf 256}, \penalty0 241--251 (Nov., 2014).

\bibitem{Murase15review}
K.~{Murase}.
  \emph{AIP Conference Series}, {\bf 1666}, p. 040006 (July,
  2015).

\bibitem{Katz13SingleSource}
B.~{Katz}, E.~{Waxman}, T.~{Thompson}, and A.~{Loeb}, arXiv:1311.0287 (Nov. 2013).

\bibitem{He13SB_HNe}
H.-N. {He}, et~al., \emph{\prd}.
  {\bf 87}\penalty0 (6):\penalty0 063011 (Mar., 2013).

\bibitem{Liu14SB_HNe}
R.-Y. {Liu}, et~al., \emph{\prd}. {\bf 89}\penalty0
  (8):\penalty0 083004 (Apr., 2014).

\bibitem{Senno15SB_HNe}
N.~{Senno}, et~al., \emph{\apj}. {\bf 806}:\penalty0 24 (June, 2015).

\bibitem{Anchordoqui14SB_nu_detection}
L.~A. {Anchordoqui}, et~al., \emph{\prd}. {\bf 89}\penalty0 (12):\penalty0
  127304 (June, 2014).

\bibitem{TQW07SB_gamma}
T.~A. {Thompson}, E.~{Quataert}, and E.~{Waxman}, \emph{\apj}. {\bf 654},
  \penalty0 219--225 (Jan., 2007).

\bibitem{Murase13100GeVconst}
K.~{Murase}, M.~{Ahlers}, and B.~C. {Lacki}, \emph{\prd}. {\bf 88}\penalty0
  (12):\penalty0 121301 (Dec., 2013).

\bibitem{Chang14SB_gbgnd}
X.-C. {Chang} and X.-Y. {Wang}, \emph{\apj}. {\bf 793}:\penalty0
  131 (Oct., 2014).

\bibitem{Tamborra14SB_gbgnd}
I.~{Tamborra}, S.~{Ando}, and K.~{Murase}, \emph{\jcap}. {\bf 9}:\penalty0 043 (Sept.,
  2014).

\bibitem{Kashti08}
T.~{Kashti} and E.~{Waxman}, \emph{JCAP}. {\bf 5},
  \penalty0 6--+ (May, 2008).

\bibitem{Auger15GZK_nu_limit}
A.~{Aab}, et~al., \emph{\prd}. {\bf 91}\penalty0
  (9):\penalty0 092008 (May, 2015).

\bibitem{Gorham13Ballon}
P.~W. {Gorham}, \emph{Nuclear Physics B Proc. Supp.} {\bf 243},
  \penalty0 231--238 (Oct., 2013).

\bibitem{Allison14ARA}
{ARA Collaboration}, P.~{Allison}, et~al., arXiv:1404.5285 (Apr. 2014).

\bibitem{Barwick15AIANNA}
S.~W. {Barwick}, et~al.,
  \emph{Astroparticle Physics}. {\bf 70}, \penalty0 12--26 (Oct., 2015).

\end{thebibliography}

\end{document}